\documentclass[a4paper,12pt]{article}

\usepackage[usenames]{color}

\usepackage{graphicx}

\usepackage{array}

\usepackage{amsmath}
\usepackage{amssymb}
\usepackage{theorem}
\usepackage{epsf}
\usepackage{latexsym}

\theoremstyle{break}

\def\therefore{.\raise1ex\hbox{.}.}
\def\because{\raise1ex\hbox{.}.\raise1ex\hbox{.}}

\title{\bf{Non-local Wess-Zumino Model on Nilpotent Noncommutative Superspace}}

\author{Yoshishige Kobayashi\thanks{E-mail : yosh@phys.metro-u.ac.jp} \ and Shin Sasaki\thanks{E-mail : shin-s@phys.metro-u.ac.jp}}

\date{\empty}

\begin{document}
\begin{titlepage}
\maketitle
\thispagestyle{empty}
\begin{center}
{\it Department of Physics, Faculty of Science,\\
     Tokyo Metropolitan University,\\
     1--1 Minami-osawa, Hachioji-shi,\\
     Tokyo 192--0397, Japan}
\end{center}

\vspace{2cm}

\begin{abstract}
We investigate the theory of the bosonic-fermionic noncommutativity,
 $[x^{\mu},\theta^{\alpha}] = i \lambda^{\mu \alpha}$, and the
 Wess-Zumino model deformed by the noncommutativity. Such noncommutativity links
 well-known space-time noncommutativity to superspace non-anticommutativity.
 The deformation has the nilpotency. We can explicitly evaluate noncommutative 
 effect in terms of new interactions between component
 fields. The interaction terms that have Grassmann 
 couplings are induced. The noncommutativity does completely break full
 $\mathcal{N}=1$ supersymmetry to $ \mathcal{N} = 0 $ theory
 in Minkowski signature. Similar to the space-time noncommutativity, this theory has
 higher derivative terms and becomes non-local theory. However this
 non-locality is milder than the space-time noncommutative field theory.
 Due to the nilpotent feature of the coupling constants, we find that there are only
 finite number of Feynman diagrams that give noncommutative corrections at each loop order.
\end{abstract}

\end{titlepage}

\section{Introduction}
The noncommutative field theory was originally suggested by Snyder \cite{Snyder} from the view point 
of regularizing ultraviolet (UV) divergence of ordinary quantum field
theory (QFT).
In recent years, quantum field theories on noncommutative (NC) space-time 
have been studied, since the relation to string theory was discovered \cite{Seiberg-Witten}.

In the string framework, noncommutativity among space-time coordinates 
\begin{equation}
[x^{\mu},x^{\nu}] = i \theta^{\mu\nu}
\end{equation}
is realized on the D-brane world volume  when there exists NS-NS B-field
backgrounds \cite{Seiberg-Witten}. 
This noncommutativity can
be represented by Moyal-Weyl star product $\star = 
\exp[\frac{i}{2} \theta^{\mu \nu} \overleftarrow{\partial}_{\mu} 
\overrightarrow{\partial}_{\nu}]$ which is used 
to construct NC QFT on ordinary (commutative) space.
NC QFT essentially contains infinite number of differential operators, 
therefore, becomes non-local theory .
The quantization of the NC field theory in terms of higher derivative theory
was studied in the literature \cite{Fujikawa}. Many interesting feature in this field has been investigated
: UV/IR mixing, noncommutative instanton
and relation to string and matrix theory, for example.

On the other hand, 
supersymmetric extension of ordinary field theories is itself very 
important for many reasons. Usually, supersymmery (SUSY) makes a theory very
tractable and sometimes completely solvable. Four-dimensional $\mathcal{N}=1$ SUSY theory is the most efficiently 
formulated by use of the superspace $z^{M} = 
(x^m,\theta^{\alpha},\bar{\theta}^{\dot{\alpha}})$ and superfields on it. 

Recently, non(anti)commutativity of fermionic coordinates 
\begin{equation}
\{\theta^{\alpha},\theta^{\beta}\} = C^{\alpha \beta}
\end{equation}
in four-dimensional Euclidean $\mathcal{N}=1$ superspace was 
considered by turning on constant Ramond-Ramond five-form field 
strength $F^{\alpha \beta}$ in Calabi-Yau compactification with 
wrapped D-brane \cite{Seiberg,Ooguri-Vafa,Seiberg-Berkovits}. 
Corresponding to this non(anti)commutativity, Seiberg considered Moyal-Weyl 
star product among superfields
\begin{equation}
f(\theta) \star g(\theta) = f(\theta) \exp\left[ -\frac{C^{\alpha}}{2} 
\frac{\overleftarrow{\partial}}{\partial \theta^{\alpha}} 
\frac{\overrightarrow{\partial}}{\partial \theta^{\beta}} \right]
g(\theta). \label{SUSY_star_product}
\end{equation}
This corresponds to Q-deformation (non supersymmetric deformation) of
SUSY field theory \cite{Ferrara}. It breaks half of original SUSY and gives $\mathcal{N} = 1/2$ SUSY in four-dimensional 
D-brane world volume.
This star product is nilpotent thanks to their Grassmann property and 
adds {\it finite} number of terms to the ordinary
Lagrangian. Because it has only fermionic differential, the additional
component terms do not contain {\it any} extra space-time derivative. This is the difference between
superspace non(anti)commutativity and space-time one.
 The latter case is inevitably non-local. Many aspects of this 
non(anti)commutative (NAC) supersymmetric field theory was studied \cite{NAC}.

In this paper, we propose another type of deformation
such as bosonic-fermionic noncommutativity
\begin{equation}
[x^{\mu},\theta^{\alpha}] = i \lambda^{\mu \alpha}.
\end{equation}
This noncommutativity is the middle point of boson-boson (space-time) 
and fermion-fermion (superspace) noncommutativity. The QFT on this space 
links usual space-time noncommutative QFT to NAC supersymmetric 
theory. The Moyal-Weyl star product corresponding to this noncommutativity 
is also nilpotent and gives {\it finite} number of corrections to
ordinary supersymmetric theories. Because of the space-time derivative
in the star product
the theory has higher derivative terms and becomes non-local, but the
effect is milder than usual NC field theory. We find that this deformation breaks all of SUSY
in Minkowski signature. A notable feature of this theory is
that new interaction has Grassmann coupling. Therefore, induced vertices have
nilpotent property in the perturbative calculus.

This paper is organized as follows. In section \ref{NAC}, we
review the NAC $\mathcal{N}=1/2$ theory in the framework of Seiberg's 
work \cite{Seiberg} which is prerequisite to understand our work. In
section \ref{BF}, we introduce our setup and give explicit form 
of the Moyal-Weyl product. In section \ref{WZ}, we show that 
noncommutative deformation of
the Wess-Zumino Lagrangian gives finite number of additional terms, 
only up to fourth order differential. 
In section \ref{Quantum}, we discuss the quantum aspects of this deformed Wess-Zumino 
model. In section \ref{string}, we discuss the geometric origin of this 
noncommutativity, section \ref{discussion} is devoted to the discussion.

\section{Non(anti)commutative $\mathcal{N}=1/2$ supersymmetric theory}\label{NAC}
In \cite{Seiberg}, Seiberg considered $\mathcal{N}=1/2$ supersymmetric theory based on the 
non(anti)commutative setup
\begin{equation}
\{\theta^{\alpha},\theta^{\beta}\} = C^{\alpha\beta}, \qquad 
 \{\bar{\theta}^{\dot{\alpha}}, \bar{\theta}^{\dot{\beta}}\} = 0.
\end{equation}
This setup is possible in the Euclidean space-time. In that case $\theta$
and $\bar{\theta}$ are independent variables. 
Corresponding to this relation, we use Moyal-Weyl star product
(\ref{SUSY_star_product}) among superfields. In \cite{Seiberg} he considered a simple situation 
such that chiral coordinates $y^{\mu}
= x^{\mu} + i \theta^{\alpha} \sigma^{\mu}_{\alpha \dot{\alpha}}
\bar{\theta}^{\dot{\alpha}}$ do commute with all of coordinates, {\it i.e.}
\begin{eqnarray}
[y^{\mu},y^{\nu}] = [y^{\mu},\theta^{\alpha}] = [y^{\mu},\bar{\theta}^{\dot{\alpha}}]
 = 0.
\end{eqnarray}
Though these relations imply space-time noncommutativity $[x^{\mu},x^{\nu}]
\not= 0$. This is because original world sheet coordinates $x^{\mu}$
are affected by the graviphoton background field (vertex operator),
while chiral coordinates $y^{\mu}$ is free from introduced graviphoton background.

In the chiral base, the Moyal-Weyl star product (\ref{SUSY_star_product}) can be written by
supercharge $Q_{\alpha}$, namely
\begin{eqnarray}
\star = \exp\left[ - \frac{C^{\alpha \beta}}{2}
	     \overleftarrow{Q}_{\alpha} \overrightarrow{Q}_{\beta}
	    \right].
\end{eqnarray}
This star product gives non-supersymmetric deformation and $\cal{N}=$1 SUSY is broken 
to $\mathcal{N} = 1/2$  due to the relation  $\{Q_{\alpha},\overline{Q}_{\dot{\beta}}\} \not= 0$.
On the other hand, by the help of the properties $\{D_{\alpha},Q_{\beta}\} = \{
\overline{D}_{\dot{\alpha}},Q_{\beta}\} = \{D_{\alpha} ,
\overline{Q}_{\dot{\beta}}\} = \{ \overline{D}_{\dot{\alpha}},
\overline{Q}_{\dot{\beta}}\} = 0$,
we can easily construct the deformed version of supersymmetric model preserving chirality. From these
facts, Seiberg showed the non(anti)commutative effect gives only one
induced interaction term to the original Wess-Zumino Lagrangian;
\begin{eqnarray}
\mathcal{L} &=& \int \! d^4 \ \theta \overline{\Phi} \star \Phi + \int \!
 d^2 \theta \ \left( \frac{1}{2} m \Phi \star \Phi + \frac{1}{3} g \Phi
	       \star \Phi \star \Phi \right) + \int \!
 d^2 \bar{\theta} \ \left( \frac{1}{2} \bar{m} \overline{\Phi} \star \overline{\Phi} + \frac{1}{3} \bar{g} \overline{\Phi}
	       \star \overline{\Phi} \star \overline{\Phi} \right)
 \nonumber \\
&=& \mathcal{L}(C=0) - \frac{1}{3} g \det C F^3 + \mathrm{(total \
 derivative)}.
\end{eqnarray}
This apparently preserves $Q$ SUSY but does not $\overline{Q}$. 
We need to consider only the deformation part to grasp the
physical meanings of this non(anti)commutative $\mathcal{N} = 1/2$
supersymmetric theory.

The $\mathcal{N} = 1/2$ supersymmetric Yang-Mills theory can also be
constructed in the same manner, {\it i.e.} by replacing each product
with the star product,
\begin{eqnarray}
W_{\alpha} &=& \left. - \frac{1}{4} \overline{D} \overline{D} e^{-V} D_{\alpha}
 e^V \right|_{\star} \nonumber , \\
 \overline{W}_{\dot{\alpha}} &=& \left. \frac{1}{4} D D e^V
 \overline{D}_{\dot{\alpha}} e^{-V} \right|_{\star}.
\end{eqnarray}
The gauge transformation is also achieved by the star product and
Seiberg found the field redefinition that allows component fields to transform in the canonical
way. The deformed Lagrangian of the gauge fields in the modified Wess-Zumino gauge is
\begin{eqnarray}
\int \! \! d^2 \theta \ \mathrm{Tr}W \star W &=& \int \! \! d^2 \theta \
 \mathrm{Tr} W W - i C^{\mu \nu} \mathrm{Tr} F_{\mu \nu} \bar{\lambda}
 \bar{\lambda} + \frac{|C|^2}{4} \mathrm{Tr} ( \bar{\lambda}
 \bar{\lambda})^2 ,\nonumber \\
\int \! \! d^2 \bar{\theta} \ \mathrm{Tr} \overline{W} \star \overline{W} &=& \int \! \! d^2 \bar{\theta} \
 \mathrm{Tr} \overline{W} \overline{W} - i C^{\mu \nu} \mathrm{Tr} F_{\mu \nu} \bar{\lambda}
 \bar{\lambda} + \frac{|C|^2}{4} \mathrm{Tr} ( \bar{\lambda}
 \bar{\lambda})^2 + \mathrm{(total \ derivative)}.\nonumber
\end{eqnarray}

\section{Bosonic-Fermionic noncommutativity}\label{BF}
In this section, we consider mixed noncommutativity
between bosonic and fermionic coordinates in the $\mathcal{N}=1$ superspace,
\begin{equation}
[x^{\mu},\theta^{\alpha}] = i \lambda^{\mu \alpha}. \label{BF_noncom}
\end{equation}
Note that the way of noncommutative deformation of the Lagrangian depends on which
signature we choose, namely, Minkowski or Euclidean.
Because in the Euclidean space we can choose $
\theta$ and $\bar{\theta}$ independently but in the Minkowski space
they are related with each other by Hermitian conjugate.  
When we are in the Minkowski space the relation
(\ref{BF_noncom}) induces $[x^{\mu},\bar{\theta}^{\dot{\alpha}}] = i
\bar{\lambda}^{\mu \dot{\alpha}}$ by Hermite conjugation.
On the other hand, the Euclidean deformation can keep 
``conjugation'' piece commutative
in spite of noncommutative relation (\ref{BF_noncom}) \footnote{The Minkowski formulations of NAC superspace can be found in
\cite{Chaichian,Nazaryan}.}. Notice that this deformation breaks $\mathcal{R}$-symmetry explicitly but gives
consistent non(anti)commutative geometry in the Minkowski space
compatible with supertranslations \cite{Dietmar}.

In the Minkowski situation we should carefully formulate the deformation of the theory.
We require that chiral coordinates $y^{\mu} = x^{\mu} + i \theta^{\alpha} 
\sigma^{\mu}_{\alpha\dot{\beta}} \bar{\theta}^{\dot{\beta}}$ to be 
commutative; $[y^{\mu},y^{\nu}] =  [y^{\mu},\bar{\theta}^{\dot{\alpha}}] = 0$,
other fermionic coordinates hold Grassmann property
 $ \{\theta^{\alpha},\theta^{\beta}\} = \{ 
\theta^{\alpha}, \bar{\theta}^{\dot{\beta}} \} = 
\{\bar{\theta}^{\dot{\beta}},\bar{\theta}^{\dot{\beta}}\} = 0$.
The corresponding Moyal-Weyl star product is written as
\begin{eqnarray}
f(y,\theta) \star g(y,\theta) &=& f(y,\theta) \exp\left[ \frac{i}{2} 
\lambda^{\mu \alpha} \left( \frac{\overleftarrow{\partial}}{\partial 
y^{\mu} } \frac{\overrightarrow{\partial}}{\partial 
\theta^{\alpha} } - \frac{\overleftarrow{\partial}}{\partial 
\theta^{\alpha} } \frac{\overrightarrow{\partial}}{\partial 
y^{\mu} }  \right) \right] g(y,\theta).
\end{eqnarray}
Our notation and convention are standard \cite{WB} and the other useful
formulae are available in appendix.
We adopt non-Hermitian star product, {\it i.e.}
$(\star)^{\dagger} = \bar{\star} \not= \star$. The $\bar{\star}$ is
formal Hermite conjugate of $\star$ product
\begin{eqnarray}
\bar{\star} = \exp\left[ \frac{i}{2} \bar{\lambda}^{\mu \dot{\alpha}}
		   \left(  \frac{\overleftarrow{\partial}}{\partial
						      \bar{y}^{\mu}}
						      \frac{\overrightarrow{\partial}}{\partial
						      \bar{\theta}^{\dot{\alpha}}} - \frac{\overleftarrow{\partial}}{\partial
						      \bar{\theta}^{\dot{\alpha}}}
						      \frac{\overrightarrow{\partial}}{\partial
						      \bar{y}^{\mu}}    \right)\right],
\end{eqnarray}
and has property $\left(f(y,\theta) \star g(y,\theta) \right)^{\dagger}
= \bar{g} (\bar{y},\bar{\theta}) \bar{\star} \bar{f}
(\bar{y},\bar{\theta})$ \cite{Ferrara}\cite{Chaichian}. We should stress
that, the star product $\star$ breaks $\bar{Q}$-SUSY but preserves
$Q$-SUSY while the conjugate star product $\bar{\star}$ breaks $Q$-SUSY
but preserves $\bar{Q}$-SUSY.

Alternative suggestion was proposed. In \cite{Nazaryan} they used
Hermitian star product $* = *^{\dagger}$. But their star product is not
associative and needs some reordering of the fields. Here, we require
associativity of product. This formulation is natural since
non-associative deformation may be related to the curved background
setup \cite{Cornalba} and in that case the effective world volume
theory is described by the Kontsevich type deformation \cite{Kontsevich}. 

Explicit expansion of our star product is
\begin{eqnarray}
\star &=& \exp\left[\frac{i}{2} \lambda^{\mu \alpha} \left(
						      \frac{\overleftarrow{\partial}}{\partial
						      y^{\mu}}
						      \frac{\overrightarrow{\partial}}{\partial
						      \theta^{\alpha}} - \frac{\overleftarrow{\partial}}{\partial
						      \theta^{\alpha}}
						      \frac{\overrightarrow{\partial}}{\partial
						      y^{\mu}}
						     \right)\right]
\nonumber \\
&=& 1 + \frac{i}{2} \lambda^{\mu \alpha} \left( \frac{\overleftarrow{\partial}}{\partial
						      y^{\mu}}
						      \frac{\overrightarrow{\partial}}{\partial
						      \theta^{\alpha}} - \frac{\overleftarrow{\partial}}{\partial
						      \theta^{\alpha}}
						      \frac{\overrightarrow{\partial}}{\partial
						      y^{\mu}} \right)
\nonumber \\
& & + \frac{1}{8} \lambda^{\mu \alpha} \lambda^{\nu \beta} \left( \frac{\overleftarrow{\partial}}{\partial
						      y^{\mu \nu}}
						      \frac{\overrightarrow{\partial}}{\partial
						      \theta^{\alpha
						      \beta}} + 2
						      \frac{\overleftarrow{\partial}}{\partial
						      y^{\mu} \partial
						      \theta^{\beta}}
						      \frac{\overrightarrow{\partial}}{\partial
						      \theta^{\alpha}
						      \partial y^{\nu}} + \frac{\overleftarrow{\partial}}{\partial
						      \theta^{\alpha \beta}}
						      \frac{\overrightarrow{\partial}}{\partial
						      y^{\nu\mu}}
								 \right)
\nonumber \\
& & + \frac{i}{16} \lambda^{\mu \alpha} \lambda^{\nu \beta}
 \lambda^{\rho \gamma} \left( \frac{\overleftarrow{\partial}}{\partial
						       y^{\rho} \partial
						       \theta^{\alpha \beta}}
						      \frac{\overrightarrow{\partial}}{\partial
						      \theta^{\gamma} \partial
						      y^{\nu\mu}} - \frac{\overleftarrow{\partial}}{\partial
						      y^{\mu\rho}
						      \partial \theta^{\beta}}
						      \frac{\overrightarrow{\partial}}{\partial
						      \theta^{\gamma\alpha}\partial
						      y^{\nu}}
		       \right) \nonumber \\
& & + \frac{1}{64} \lambda^{\mu \alpha} \lambda^{\nu \beta}
 \lambda^{\rho \gamma} \lambda^{\sigma \delta} \frac{\overleftarrow{\partial}}{\partial
						      y^{\mu\rho}\partial
						      \theta^{\beta \delta}}
						      \frac{\overrightarrow{\partial}}{\partial
						      \theta^{\gamma
						      \alpha} \partial
						      y^{\sigma \nu}} . \label{explicit_star}
\end{eqnarray}
Here we denote second derivatives as $ \frac{\overrightarrow{\partial}^2}{\partial y^{\mu}
\partial y^{\nu}} \equiv \frac{\overrightarrow{\partial}}{\partial y^{\mu \nu}}$ and $
\frac{\overrightarrow{\partial^2}}{\partial \theta^{\alpha} \partial \theta^{\beta}} 
\equiv \frac{\overrightarrow{\partial}}{\partial \theta^{\alpha
\beta}}$. From this we see the relation
\begin{eqnarray}
& &[y^{\mu},\theta^{\alpha}]_{\star} = i \lambda^{\mu \alpha} , \quad 
[y^{\mu},\bar{\theta}^{\dot{\alpha}}]_{\star} = 0, \quad
[\bar{y}^{\mu},\theta^{\alpha}]_{\star} = i \lambda^{\mu \alpha}, \nonumber \\
& & [y^{\mu},y^{\nu}]_{\star} = 0, \quad [\bar{y}^{\mu} ,
 \bar{\theta}^{\dot{\alpha}}]_{\star} = 0,
\end{eqnarray}
and
\begin{eqnarray}
& &[y^{\mu},\bar{y}^{\nu}]_{\star} = 2 \lambda^{\mu\alpha} 
(\sigma^{\nu})_{\alpha \dot{\beta}} \bar{\theta}^{\dot{\beta}}, \nonumber \\
& &[x^{\mu},x^{\nu}]_{\star} = \lambda^{\mu\alpha} 
(\sigma^{\nu})_{\alpha \dot{\beta}} \bar{\theta}^{\dot{\beta}} - 
\lambda^{\nu\alpha} (\sigma^{\mu})_{\alpha \dot{\beta}} \bar{\theta}^{\dot{\beta}}.
\end{eqnarray}
The Hermitian conjugate of these relations leads to consistent results --- for
example $[\bar{y}^{\mu} , \bar{\theta}^{\dot{\alpha}}]_{\bar{\star}} = i
\bar{\lambda}^{\mu \dot{\alpha}}$.

Contrary to the Minkowski case, in our case we need only one star product $\star$ to
formulate the deformation theory in the Euclidean case because we don't
require the Hermite Lagrangian.

\section{Wess-Zumino model on the noncommutative superspace}\label{WZ}
We can treat the chiral superfields appropriately,
because our star product is the Q-deformation which preserves chirality, 
{\it i.e.} the products of chiral superfields are again chiral superfields.
The chiral superfield $\Phi$ is defined by the ordinary condition
$\overline{D}_{\dot{\alpha}} \Phi = 0$ and the aiti-chiral superfield $\overline{\Phi}$ is
$D_{\alpha} \overline{\Phi} = 0$. The general solution of
this constraint for the chiral superfield is the function of $y^m$ and
$\theta$,
\begin{eqnarray}
\Phi (y,\theta) = A(y) + \sqrt{2} \theta^{\alpha} \psi_{\alpha} (y) +
 \theta^2 F(y). \label{chiral_expansion}
\end{eqnarray}
One may consider that equation
(\ref{chiral_expansion}) should be expanded by using the star product. But in our case the star product doesn't introduce $\bar{\theta}$
so we can redefine the component fields after explicit use of equation
(\ref{explicit_star}).
The anti-chiral superfield expansion is Hermitian conjugate of the
chiral superfield expansion (\ref{chiral_expansion}) in the Minkowski space.
\subsection{Lagrangian in the Euclidean space}
In the Euclidean space, we can consider the noncommutative relation only in
the chiral sector
\begin{eqnarray}
& & [y^{\mu},\theta^{\alpha}] = i \lambda^{\mu \alpha} , \nonumber \\
& & [ \bar{y}^{\mu} , \bar{\theta}^{\dot{\alpha}} ] = 0.
\end{eqnarray}
The Wess-Zumino action which is deformed on this noncommutative superspace is
\begin{eqnarray}
\mathcal{L} = \int \!\! d^4 \theta \ \overline{\Phi} \star \Phi  + \int \! \! d^2 \theta \ \left( \frac{m}{2} \Phi \star \Phi +
			       \frac{g}{3} \Phi \star \Phi \star \Phi
			      \right) + \int \! \! d^2 \bar{\theta} \ \left( \frac{\bar{m}}{2} \overline{\Phi} \star \overline{\Phi} +
			       \frac{\bar{g}}{3} \overline{\Phi} \star
			       \overline{\Phi} \star \overline{\Phi}
								      \right)
\end{eqnarray}
After expanding the Moyal-Weyl products and doing partial integrations, we
find the component action of the deformed theory;
\begin{eqnarray}
S &=& \int \!\! d^4 x \mathcal{L}_0 + \frac{i}{\sqrt{2}} \lambda^{\mu
 \alpha} \psi_{\alpha} ( \partial_{\mu} \Box \bar{A}) + \frac{1}{\sqrt{2}}
 \lambda^{\mu \alpha} F ( \sigma^{\nu})_{\alpha \dot{\alpha}}
( \partial_{\mu} \partial_{\nu} \bar{\psi}^{\dot{\alpha}})  - \frac{1}{2}
 \lambda^{\mu \nu} F ( \partial_{\mu} \partial_{\nu} \Box \bar{A})
 \nonumber \\
& & + \frac{g}{2} \lambda^{\mu \alpha} \lambda^{\nu 
\beta} F ( \partial_{\mu} \psi_{\beta}) ( \partial_{\nu} \psi_{\alpha}) + 
g \lambda^{\mu \nu} F ( \partial_{\mu} F) ( \partial_{\nu} A ) + 
\frac{g}{3} \lambda^{\mu \nu \rho \sigma} F ( \partial_{\mu} \partial_{\nu} 
F) ( \partial_{\rho} \partial_{\sigma}  F )\nonumber \\
& & + \frac{\bar{g}}{3} \lambda^{\mu
 \nu} \left[ - ( \partial_{\mu} \bar{A})
					 ( \partial_{\nu} \bar{A})
					 ( \Box \bar{A}) +
					 ( \partial_{\rho} \bar{A})
					 ( \partial^{\rho} \bar{A})
					 ( \partial_{\mu} \partial_{\nu}
					  \bar{A}) \right].
\end{eqnarray}
Here $\mathcal{L}_0$ is the undeformed (original) Wess-Zumino
Lagrangian. Because of the spacetime-superspace-mixed noncommutativity,
the higher derivative terms are induced.
In this Euclidean setup the deformation breaks Hermicity. In the next
section we will show that Hermite deformation can be 
obtained in the Minkowski space. 
\subsection{Lagrangian in the Minkowski space}
Noncommutative deformed Minkowski Wess-Zumino model is also obtained by
replacing the ordinary products with the Moyal-Weyl star products.
But in this case we should be careful because we want to keep the
theory Hermite. 

The deformation of the interaction parts is straightforward. We can simply replace ordinary 
product with star
\begin{eqnarray}
\frac{m}{2} \Phi \Phi + \frac{g}{3} \Phi \Phi \Phi + \mathrm{(h.c.)} 
\longrightarrow \frac{m}{2} \Phi \star \Phi + \frac{g}{3} 
\Phi \star \Phi \star \Phi + \mathrm{(h.c.)}.
\end{eqnarray}
Here (h.c.) in the RHS represents Hermitian conjugated parts, namely, $ \frac{\bar{m}}{2}
\overline{\Phi} \bar{\star} \overline{\Phi} + \frac{\bar{g}}{3} 
\overline{\Phi} \bar{\star} \overline{\Phi} \bar{\star} \overline{\Phi}$. 

The kinetic term is little bit different. Since such term contains both
chiral and anti-chiral superfield. We simply choose the kinetic term as
$ \frac{1}{2} \overline{\Phi}
\star \Phi + \frac{1}{2} \overline{\Phi}
\bar{\star} \Phi$. Due to the property $\left( 
\overline{\Phi} \star \Phi \right)^{\dagger} = \overline{\Phi} 
\bar{\star} \Phi$, the kinetic term is Hermite \footnote{But this is not necessary because free part of 
Lagrangian is not affected by noncommutativity in the integral by 
general property of star product. Here, we define component expansion of
anti-chiral superfield as formal complex conjugate of chiral one.}.
The resulting deformed Wess-Zumino action is
\begin{eqnarray}
\mathcal{L} &=& \frac{1}{2} \int \! d^4 \theta \ \overline{\Phi} 
\star \Phi + \frac{1}{2} \int \! d^4 \theta \ \overline{\Phi} \bar{\star} 
\Phi \nonumber \\
& & \quad + \int \! d^2 \theta \left( \frac{m}{2} \Phi \star \Phi + 
\frac{g}{3} \Phi \star \Phi \star \Phi \right)  + \int \! d^2 \bar{\theta} \left( \frac{\bar{m}}{2} 
\overline{\Phi} \bar{\star} \overline{\Phi} + \frac{\bar{g}}{3} \overline{\Phi} \bar{\star} \overline{\Phi} 
\bar{\star} \overline{\Phi}\right).
\end{eqnarray}
The explicit calculation of the star product and the Grassmann
integration show the component Lagrangian of the deformed Wess-Zumino model;
\begin{eqnarray}
S &=& \int \! d^4 x \ \mathcal{L}_0  + \frac{g}{3}\int \! d^4 x \ \left[ \lambda^{\mu 
\alpha} \lambda^{\nu \beta} \left( \psi_{\alpha} ( \partial_{\mu} 
\psi_{\beta}) ( \partial_{\nu} F) + \frac{1}{2} F (\partial_{\mu} \psi_{\beta}) 
( \partial_{\nu} \psi_{\alpha}) \right) \right. \nonumber \\
& & + \lambda^{\mu \nu} \left(  F \psi^{\alpha} 
( \partial_{\mu} \partial_{\nu} \psi_{\alpha}) - A F ( \partial_{\mu} 
\partial_{\nu} F) - A ( \partial_{\mu} F ) (\partial_{\nu} F) - F^2 
( \partial_{\mu} \partial_{\nu} A) \right) \nonumber \\
& & \left. + \lambda^{\mu \nu \rho \sigma} F ( \partial_{\mu} 
\partial_{\nu} F) ( \partial_{\rho} \partial_{\sigma} F )
    \right. \bigg]+ \mathrm{[(h.c.)]} . \label{deformed_WZ_Lagrangian}
\end{eqnarray}
Here $\mathcal{L}_0$ is the undeformed original Wess-Zumino action in
the Minkowski space;
\begin{eqnarray}
 \mathcal{L}_0 &=& \bar{A} \Box A + i \partial_{\mu}
 \bar{\psi}_{\dot{\alpha}} ( \bar{\sigma}^{\mu})^{\dot{\alpha} \beta}
 \psi_{\beta} + \bar{F}F \nonumber \\
& & + \left[ m ( AF - \frac{1}{2} \psi \psi) + g ( A^2 F - \psi \psi A)
       + \mathrm{(h.c.)}  \right].
\end{eqnarray}
The generalized (including higher derivative terms) equation of motion
\cite{Gupta} can be applied. The equations of motion of the auxiliary fields are
\begin{eqnarray}
F &=& -(\bar{m} \bar{A} + \bar{g} \bar{A}^{2}) \nonumber \\
& & + \bar{g} \bar{\lambda}^{\mu \nu} \bar{F}
\partial_{\mu} \partial_{\nu} \bar{A} -
\frac{1}{2} \bar{g} \bar{\lambda}^{\mu \dot{\alpha}}
 \bar{\lambda}^{\nu \dot{\beta}}
(\partial_{\mu} \bar{\psi}_{\dot{\beta}}
 \partial_{\nu} \bar{\psi}_{\dot{\alpha}} ) \nonumber \\
& &- \frac{1}{3} \bar{g} \bar{\lambda}^{\mu \nu \rho \sigma}
\left\{ (\partial_{\mu} \partial_{\nu} \bar{F} )(\partial_{\rho}
\partial_{\sigma} \bar{F} )
+ 2 \partial_{\mu \nu}( \bar{F} \partial_{\rho \sigma} \bar{F} )
\right\}, \label{EOM_Fbar}.
\end{eqnarray}
and Hermitian conjugate one.

Eq.(\ref{EOM_Fbar}) looks unusual because
the auxiliary field $\bar{F}$ couples to other components 
or even to the derivatives of itself.
But in fact eq.(\ref{EOM_Fbar}) can be solved for $F$ and $\bar{F}$ exactly as follows:
Using the conjugate equation of (\ref{EOM_Fbar}) to eliminate $\bar{F}$ in eq.(\ref{EOM_Fbar}),
one obtains the expression of $F$ in terms of other dynamical fields
and $F$ itself.
After that, replacing $F$ in the RHS of (\ref{EOM_Fbar}) with the
expression of the whole the RHS of $F$  -- like the successive approximation.
This recursive replacement is done iteratively but    
will stop thanks to the nilpotency of couplings $\lambda$,
since every $F$ in the RHS appears with $\lambda$.
In the end one get $F$ written by other dynamical components only.
It is not difficult to see that no auxiliary fields
become dynamical by the deformation, in other words, new degrees of freedom on shell never appear.

Let us check the SUSY transformation of the action. The undeformed part
is invariant under the $Q$-supersymmetry transformation
\begin{eqnarray}
& & \delta_{\xi} A = \sqrt{2} \xi \psi, \nonumber \\
& & \delta_{\xi} \psi = \sqrt{2} \xi F, \nonumber \\
& & \delta_{\xi} F = 0.
\end{eqnarray}
The $\lambda$-dependent parts $\mathcal{L}_{\lambda}$ in (\ref{deformed_WZ_Lagrangian}) is also invariant under the $Q$-supersymmetry transformation
\begin{eqnarray}
\frac{3}{g} \delta \mathcal{L}_{\lambda} &=& \sqrt{2} \lambda^{\mu \alpha}
 \lambda^{\nu \beta} \left( \xi_{\alpha} F \partial_{\mu} \psi_{\beta}
		      \partial_{\nu} F + \psi_{\alpha} \partial_{\mu}
		      \xi_{\beta} F \partial_{\nu} F + \frac{1}{2} F
		      \partial_{\mu} \xi_{\beta} F \partial_{\nu}
		      \psi_{\alpha} + \frac{1}{2} F \partial_{\mu}
		      \psi_{\beta} \partial_{\nu} \xi_{\alpha} F 
		     \right) \nonumber \\
& & - \sqrt{2} \lambda^{\mu \nu} \left( - F \xi^{\alpha} F
					    \partial_{\mu}
					    \partial_{\nu} \psi_{\alpha}
					    - F \psi^{\alpha}
					    \partial_{\mu}
					    \partial_{\nu} \xi_{\alpha}
					    F + \xi^{\alpha}
					    \psi_{\alpha} F
					    \partial_{\mu}
					    \partial_{\nu} F
					   \right. \nonumber \\
& & \qquad \qquad \qquad \left. + \xi^{\alpha} \psi_{\alpha} \partial_{\mu}
		       F \partial_{\nu} F + F^2 \partial_{\mu}
		       \partial_{\nu} \xi^{\alpha} \psi_{\alpha} \right)
\nonumber \\
&=& \sqrt{2} \lambda^{\mu \alpha} \lambda^{\nu \beta} \psi_{\alpha}
 \xi_{\beta} \partial_{\mu} F \partial_{\nu} F - \sqrt{2}
 \lambda^{\mu \nu} \xi^{\alpha} \psi_{\alpha} \partial_{\mu} F
 \partial_{\nu} F \nonumber \\
&=& 0.
\end{eqnarray}
We used the identity $ \lambda^{\mu \alpha} \partial_{\mu} F
\lambda^{\nu \beta} \partial_{\nu} F = - \varepsilon^{\alpha \beta} \lambda^{\mu \nu} \partial_{\mu} F
\partial_{\nu} F $. But $\mathcal{L}_{\lambda} $ is not invariant under
the $\bar{Q}$-supersymmetry. The $\bar{\lambda}$-dependent parts is
invariant under the $\bar{Q}$-supersymmetry but not $Q$-supersymmetry.
Then all of the supersymmetry is broken as a
whole. We notice that there is remaining symmetry
\begin{eqnarray}
& & \delta_{\xi} A = \sqrt{2} \xi \psi ,\quad  \delta_{\bar{\xi}} \bar{A} =
 \sqrt{2} \bar{\xi} \bar{\psi} , \nonumber \\
& & \delta_{\xi} \psi = \sqrt{2} \xi F ,\quad  \delta_{\bar{\xi}} \bar{\psi} =
 \sqrt{2} \bar{\xi} \bar{F} ,\nonumber \\
& & \delta_{\xi} F = 0 ,\quad  \delta_{\bar{\xi}} \bar{F} = 0.
\end{eqnarray}
\section{Quantum property of the deformed Wess-Zumino model}\label{Quantum}
In this section, we consider quantum aspects of the deformed Wess-Zumino
model in the Minkowski space. As we mentioned in the previous section,
$\mathcal{N} = 1$ supersymmetry is completely broken to $\mathcal{N} =
0$ in the Minkowski space. Therefore we use the component formalism
rather than the supergraph to study quantum effects.
The component formalism of the Wess-Zumino model was studied in the literature \cite{Wess-Zumino,Iliopoulos}.

The non-local terms which are induced by the deformation have at least one Grassmann (even) coupling
constant. These terms give the new vertices in the Feynman diagrams but
their additional contributions to the quantum effects are nilpotent. 

Let us separate the Lagrangian into the two parts, namely, undeformed part
$\mathcal{L}_0$ and the induced part $\mathcal{L}_{\lambda}$. To
simplify the quantum calculation, we do partial integrations and
dropping total derivative terms. We get the $\lambda (\bar{\lambda})$-dependent interaction terms;
\begin{eqnarray}
\!\!\! \mathcal{L}_{\lambda} \!\!\! &\equiv& \!\!\! \frac{g}{2} \lambda^{\mu \alpha} \lambda^{\nu 
\beta} F ( \partial_{\mu} \psi_{\beta}) ( \partial_{\nu} \psi_{\alpha}) + 
g \lambda^{\mu \nu} F ( \partial_{\mu} F)( \partial_{\nu} A) + 
\frac{g}{3} \lambda^{\mu \nu \rho \sigma} F ( \partial_{\mu} \partial_{\nu} 
F) ( \partial_{\rho} \partial_{\sigma} F) + \mathrm{(h.c.)}. \label{induced}
\end{eqnarray}

\begin{figure}[htb]
\begin{center}
\includegraphics[scale=.85]{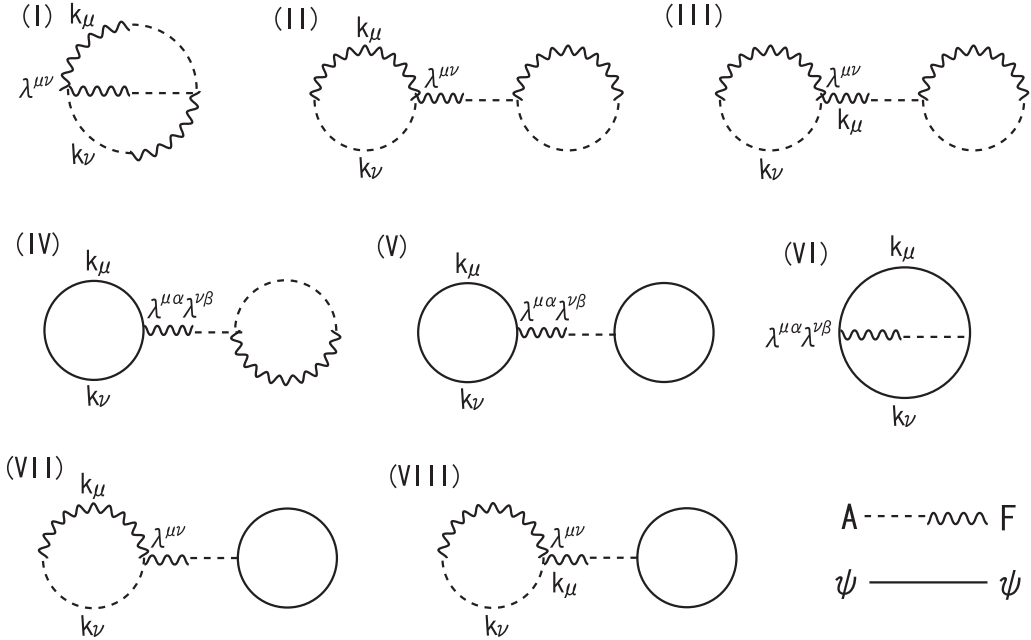}
\end{center}
\caption{Additional contribution to two loop vacuum diagram at $\mathcal{O}(\lambda^2)$}
\label{two-loop_vac}
\end{figure}

We can write the generating functional following the standard procedure, 
\begin{eqnarray}
Z[J,\eta] &=& N \exp\left[ i \int \!\! d^4x \ \mathcal{L}_{\mathrm{int}}
		  \left( \delta/i\delta J(x), \delta/\delta \eta (x)
		  \right) \right] \int \!\! \mathcal{D}\phi \exp\left[
								  i \int 
								  \!\!
								  d^4x  \left(
								  \mathcal{L}_{\mathrm{free}} + \mathcal{L}_{\mathrm{source}} \right) \right]. 
\end{eqnarray}
Here $N$ is normalization constant, $\mathcal{L}_{\mathrm{free}}$ and $\mathcal{L}_{\mathrm{int}}$ 
are free and interaction part of the Lagrangian respectively:
\begin{eqnarray}
\mathcal{L}_{\mathrm{int}} &=& \left[ g (A^2 F - \psi \psi A) \right. \nonumber \\
& & \left. + \frac{g}{2}
 \lambda^{\mu \alpha} \lambda^{\nu \beta} F \partial_{\mu} \psi_{\beta}
 \partial_{\nu} \psi_{\alpha} + g \lambda^{\mu \nu} F \partial_{\mu} F
 \partial_{\nu} A + \frac{g}{3} \lambda^{\mu \nu \rho \sigma} F
 \partial_{\mu} \partial_{\nu} F \partial_{\rho} \partial_{\sigma} F
    \right] +
 \left[ (\mathrm{h.c.}) \right].
\end{eqnarray}
$ \mathcal{L}_{\mathrm{source}}$ is the source terms
\begin{eqnarray}
\mathcal{L}_{\mathrm{source}} &=& J_A A + \bar{J}_{\bar{A}} \bar{A} + J_F F +
 \bar{J}_{\bar{F}} \bar{F} + \bar{\eta}^{\alpha} \psi_{\alpha} +
 \eta_{\dot{\alpha}} \bar{\psi}^{\dot{\alpha}}.
\end{eqnarray}
Changing the variables of path integral and using some technical
identities we find
\begin{eqnarray}
Z[J,\eta] &=& N' \exp\left[ i \int \!\! d^4x \mathcal{L}_{\mathrm{int}} \left( \delta/i\delta J(x), \delta/\delta \eta (x)
		  \right)
		       \right] \exp\left[ -i \bar{J}_{\bar{A}} \cdot
		       \Delta_{\mathrm{F}} J_A + i \bar{m} J_F \cdot
		       \Delta_{\mathrm{F}} J_A  + i m \bar{J}_{\bar{A}}
		       \cdot \Delta_{\mathrm{F}} \bar{J}_{\bar{F}}
		       \frac{}{}
				   \right. \nonumber \\
& & \left. - i	J_F \cdot \Box \Delta_{\mathrm{F}}
		       \bar{J}_{\bar{F}} - \frac{i}{2} m
		       \eta_{\dot{\alpha}} \cdot \Delta_{\mathrm{F}}
		       \eta^{\dot{\alpha}} - \frac{i}{2} \bar{m}
		       \bar{\eta}^{\alpha} \cdot \Delta_{\mathrm{F}}
		       \bar{\eta}_{\alpha} + \bar{\eta}^{\alpha} \cdot
		       \partial_{\mu} \Delta_{\mathrm{F}} (
		       \sigma^{\mu})_{\alpha \dot{\alpha}}
		       \eta^{\dot{\alpha}} \right],
\end{eqnarray}
dot stands for space-time integration. $\Delta_{\mathrm{F}}$ is 
an ordinary inverse operator
\begin{eqnarray}
\Delta_{\mathrm{F}} = \frac{1}{\Box - |m|^{2} }.
\end{eqnarray}
It should be emphasized that only {\it finite} number of 
additional vertices appear in the diagrams because of the
nilpotency. This fact enables one to get the exact expansion of the
exponential of $\mathcal{L}_{\lambda}$. The finite number of the
corrections to the diagrams at each order of perturbation appear.
To see this explicitly we expand the
exponential of $\mathcal{L}_{\lambda}$ part,
\begin{eqnarray}
Z[J,\eta] &=& N' \exp\left[ i \int \!\! d^4 x
		      \mathcal{L}_{\mathrm{int}}^{\lambda = 0} \left(
								\delta/\delta J, \delta/\delta \eta \right) \right] \left[ 1 + \int \!\! d^4 x \ \mathcal{L}_{\mathrm{int}}^{\lambda} \left( \delta/\delta J , \delta/\delta \eta  \right) + \mathrm{(terms \ up \  to \ 
\mathcal{O} (\lambda^8,\bar{\lambda}^8) )} \right]  \nonumber \\
& & \times \exp\bigg[ -i \bar{J}_{\bar{A}} \cdot
		       \Delta_{\mathrm{F}} J_A + i \bar{m} J_F \cdot
		       \Delta_{\mathrm{F}} J_A  + i m \bar{J}_{\bar{A}}
		       \cdot \Delta_{\mathrm{F}} \bar{J}_{\bar{F}}
				    \nonumber \\
& & \left. - i	J_F \cdot \Box \Delta_{\mathrm{F}}
		       \bar{J}_{\bar{F}} - \frac{i}{2} m
		       \eta_{\dot{\alpha}} \cdot \Delta_{\mathrm{F}}
		       \eta^{\dot{\alpha}} - \frac{i}{2} \bar{m}
		       \bar{\eta}^{\alpha} \cdot \Delta_{\mathrm{F}}
		       \bar{\eta}_{\alpha} + \bar{\eta}^{\alpha} \cdot
		       \partial_{\mu} \Delta_{\mathrm{F}} (
		       \sigma^{\mu})_{\alpha \dot{\alpha}}
		       \eta^{\dot{\alpha}} \right].
\end{eqnarray}
We consider the vacuum energy to grasp some quantum properties of this
model. One loop order contributions vanish due to their original
$\mathcal{N} = 1$ supersymmetry. This fact agrees with the planar and non-planar superfield calculation\footnote{There
is only planar contribution to the vacuum diagram at one-loop
order. This gives exactly zero contribution because of $\mathcal{N} = 1$
SUSY property of the original Wess-Zumino model.}. The first non-trivial
contributions come from two-loop order. Now, for simplicity, we are
going to consider only the $\mathcal{O} (\lambda^2,g^2)$ terms. There are eight
 additional Feynman diagrams [fig.\ref{two-loop_vac}].
Explicit calculation shows that all contributions cancel out exactly
at least at this order, like in the $\mathcal{N}= 1/2$ case
\cite{Terashima} in spite of the completely broken SUSY nature of this
model. Contrary to the $\mathcal{N}=1/2$ model, we have no mechanism which guarantee such cancellations.
This is very suggestive fact. The divergence structure of this model may
be milder than the usual non-SUSY models. 

\begin{figure}[htb]
\begin{center}
\includegraphics[scale=.85]{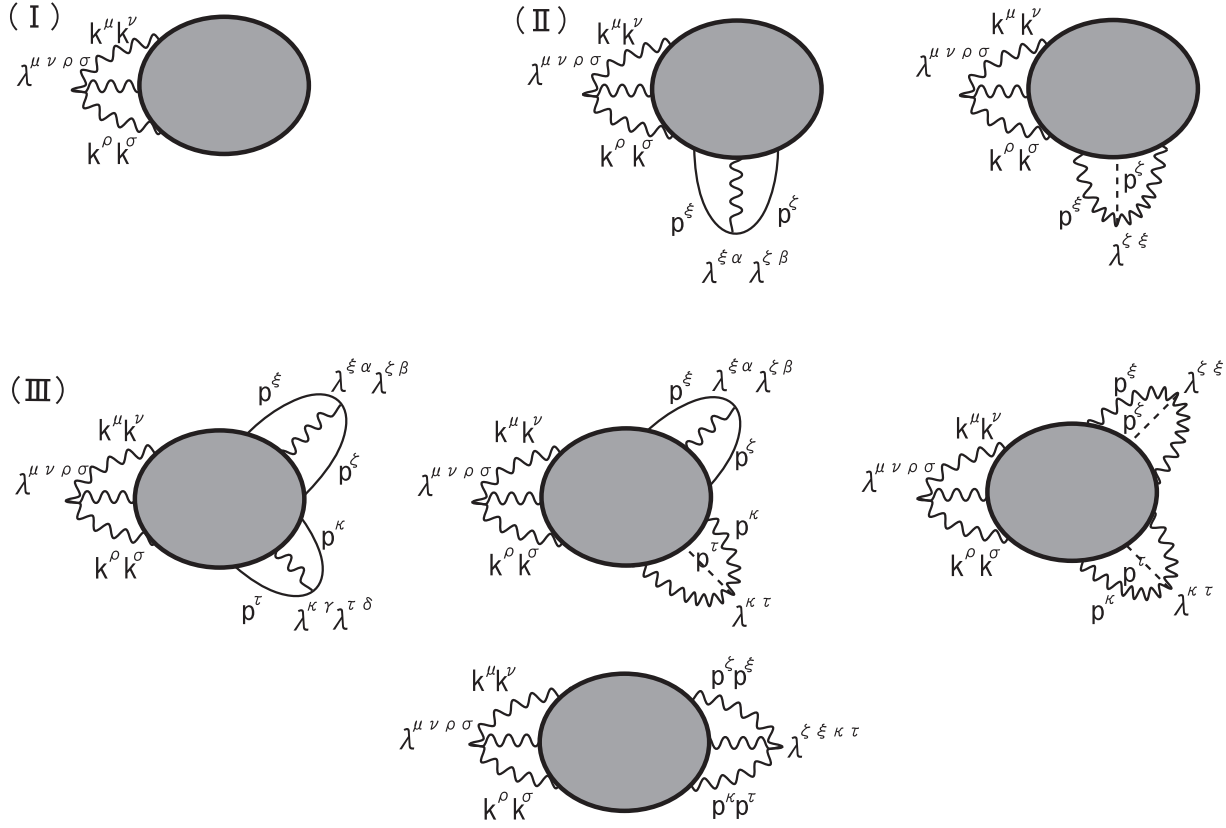}
\end{center}
\caption{F cubic contribution to the diagram. Diagram (I) is
 $\mathcal{O}(\lambda^4)$, (II) is $\mathcal{O}(\lambda^6) $ and (III) is
 $ \mathcal{O}(\lambda^8)$ contributions.}
\label{FFF}
\end{figure}

We stress that the nilpotent property of additional Feynman diagrams help us
to investigate the detail structures of higher order quantum effects. For example, when we
focus on only $\lambda$ sector (not $\bar{\lambda}$), and consider the
contributions from $\lambda^{\mu \nu \rho \sigma} F ( \partial_{\mu}
\partial_{\nu} F ) ( \partial_{\rho} \partial_{\sigma} F ) $ term, the
diagrams which contain at least one $\lambda^{\mu \nu \rho
\sigma}$ vertex are depicted in [fig.\ref{FFF}].
The shaded regions are
$\lambda=0$ contributions to the diagrams which are
already known from the $\mathcal{N}=1$ Wess-Zumino model.
We need to evaluate only finite number (seven in this case) of the diagrams to study the deformation effect. 

\section{Geometric origin}\label{string}
It is well known that NS-NS Kalb-Ramond field $B_{\mu \nu}$ induces 
the space-time noncommutativity on the open string endpoint, {\it i.e.} on the 
D-brane world volume. Recently, Ooguri and Vafa showed that when we
consider the superstring in the presence
of the graviphoton (R-R) background with the compactified six dimensional Calabi-Yau manifold, we can realize 
four-dimensional $\mathcal{N}= 1/2$ NAC superspace on the string world sheet boundary \cite{Ooguri-Vafa}. More clear 
derivation can be seen in \cite{Seiberg}. On the other hand, in \cite{Nieuwenhuizen}
they showed the generalized non(anti)commutativities among
supercoordinates in the ten dimension by considering the superstring in the constant supergravity 
backgrounds. They realized the bosonic-fermionic noncommutativity $[x^{\mu}, 
\theta^{\alpha}] = i \lambda^{\mu \alpha}$ on the open string endpoint
by turning on the constant gravitino background $\psi^{\mu}_{\alpha}$. 

The backgrounds have to be consistent with the supergravity equations of motion.
Consider the type IIB supergravity action. For simplicity we keep the
R-R field strength, dilaton $\phi$ and dilatino $\psi$ to be zero. We
take here the constant NS-NS B-field and constant gravitino. The Einstein equation 
for IIB supergravity is
\begin{equation}
R^{\mu \nu} - \frac{1}{2} g^{\mu \nu} R =T_{\mathrm{gravitino}}^{\mu\nu}.
\end{equation}
Here $H=dB$ is zero for the constant B-field and gives no
gravitational back-reaction. Taking flat limit in both sides in the Einstein equation
gives the consistent zero limit because the stress tensor of the constant spin 3/2 Rarita-Schwinger
field on the flat space vanishes\footnote{But this result is acceptable
up to quadratic order of $\psi^{\mu}_{\alpha}$ because no complete actions of the IIB
supergravity are found.}. Then the constant gravitino allows flat metric, so we can 
consider appropriate setup for the quantized superstring.

The OPE between $x$ and $\theta$ shows
\begin{eqnarray}
& &[x^{\mu},x^{\nu}] =  2 \pi i \alpha' \left( \frac{1}{g + 2 \pi
					 \alpha' B}\right)^{\mu \nu}_{A}, \nonumber \\
& &[x^{\mu}, \theta^{\alpha}] = - i (2 \pi \alpha')^2 \left( \frac{1}{g + 2 \pi
					      \alpha' B}
					     \right)^{\mu \nu}_{A} \psi^{\alpha}_{\nu} , \nonumber \\
& &\{\theta^{\alpha},\theta^{\beta}\} =  i ( 2 \pi \alpha')^3 \psi^{\alpha}_{\mu} \left(
							     \frac{1}{g
							     + 2 \pi
							     \alpha' B}
							    \right)^{\mu
\nu}_{A} \psi^{\beta}_{\nu} . \nonumber \\
\end{eqnarray}
Here we suppose the boundary condition $\theta^{\alpha} =
\tilde{\theta}^{\alpha} $ and $ \psi^{\alpha}_m = \tilde{\psi}^{\alpha}_m$.
Next we take zero-slope limit $\alpha' \to 0$ to decouple massive
mode and obtain the effective field theory. It is not possible that only $x$-$\theta$
commutator have non-zero value by the scaling itself. Note that we have to
consider the special configurations of the fields to cancel the
gravitino by the NS-NS sector in the $\theta$-$\theta$ anti-commutator.
At the same time to keep the $x$-$x$ commutator zero, we need
to scale the fields carefully. Our scaling is $\alpha' \to 0$
with $ (g+2 \pi \alpha' B)^{-1}_A = \mathrm{constant}$ and $ \psi \sim
{\alpha'}^{-2}$. Then we can realize the appropriate condition for our purpose on the
D3-brane world volume. But to reduce the supersymmetry needs further investigation.

\section{Discussion}\label{discussion}
In this paper, we studied bosonic-fermionic noncommutative superspace
and the field theory on it. There are two possibilities to deform the
theory, namely, Euclidean or Minkowski. In case of the theory in Minkowski signature, we have
to adopt non-Hermitian Moyal star product to keep the associativity, and the
resulting higher derivative structure of space-time gives non-local nature in the
theory. Our deformation is non-local in
space-time as in the bosonic (space-time) noncommutative field 
theory, and the deformation is nilpotent and additional terms to the
ordinary Lagrangian have Grassmann coupling constant. Due to this
nilpotency, there are only finite number of additional Feynman diagrams
at each loop order, and we found that the vacuum energy vanishes at the two loop
$\mathcal{O}(\lambda^2)$ in spite of the violation of
SUSY. This unexpected result tempt us to think quantum effect
is more controllable than in $\mathcal{N}=0$ theory, though further
investigations are required.

The fact that we have to
introduce {\it two} types of star products $\star$ and $\bar{\star}$ in
the Lagrangian in case of Minkowski signature gives technical
difficulties to construct the gauge theories. We have to
extract Hermitian sector from the gauge kinetic term and gauge
transformation. On the other hand in case of the Euclidean signature, we may use only
one star product by virtue of the relation $[y^{\mu},\theta^{\alpha}] = i \lambda^{\mu \alpha}$ with $[\bar{y}^{\mu} ,
\bar{\theta}^{\dot{\alpha}}] = 0 $. This case is more closely related to the Seiberg's Euclidean $\mathcal{N} = 1/2$
theory\footnote{Actually, this deformation breaks half of SUSY.}. In
this case, it is easy to construct gauge theory on the
$[y^{\mu},\theta^{\alpha}] = i \lambda^{\mu \alpha}$ space because there
are only one deformation of product $\star$ which has to be used to
define noncommutative gauge transformation.


\section*{Acknowledgments}

We would like to thank Y.~Tanii and N.~Kitazawa for useful comments. We
especially would like to thank S.~Saito for discussion about non-locality. 


\begin{appendix}
\section{Notation}
We follow the standard notation and convention \cite{WB} in both
Euclidean and Minkowski space. The fermionic differentiations are
\begin{eqnarray}
& & \frac{\overrightarrow{\partial}}{\partial \theta^{\alpha}}
 \theta^{\beta} = \delta_{\alpha}^{\beta} , \quad \theta^{\beta}
 \frac{\overleftarrow{\partial}}{\partial \theta^{\alpha}} = -
 \delta_{\alpha}^{\beta}, \nonumber \\
& & \frac{\overrightarrow{\partial}}{\partial \theta^{\alpha}}
 \theta^2 = 2 \theta_{\alpha} , \quad  \theta^2
 \frac{\overleftarrow{\partial}}{\partial \theta^{\alpha}} = 2
 \theta_{\alpha}, \nonumber \\
& & \frac{\overrightarrow{\partial}}{\partial \theta^{\alpha\beta}}
 \theta^2 = 2 \varepsilon_{\beta \alpha} , \quad \theta^2 
 \frac{\overleftarrow{\partial}}{\partial \theta^{\alpha \beta}} = - 2
 \varepsilon_{\alpha \beta},
\end{eqnarray}
and the deformation parameters are defined as follows;
\begin{eqnarray}
& & \lambda^{\mu \nu} \equiv \frac{1}{2} \lambda^{\mu \alpha}
 \lambda^{\nu}_{\ \alpha}, \qquad \bar{\lambda}^{\mu \nu} \equiv
 (\lambda^{\mu})^{\dagger} =  - \frac{1}{2} \bar{\lambda}^{\mu \dot{\alpha}}
 \bar{\lambda}^{\nu \dot{\beta}} \varepsilon_{\dot{\alpha} \dot{\beta}},
 \nonumber \\
& & \lambda^{\mu \nu \rho \sigma} \equiv \frac{1}{4} \lambda^{\mu\nu}
 \lambda^{\rho\sigma} =
 \frac{1}{16} \lambda^{\mu \alpha} \lambda^{\nu}_{\ \alpha} \lambda^{\rho
  \beta} 
 \lambda^{ \sigma}_{\ \beta} , \qquad \bar{\lambda}^{\mu \nu \rho
 \sigma} \equiv (\lambda^{\mu \nu \rho \sigma})^{\dagger} = \frac{1}{4}
 \bar{\lambda}^{\mu \nu} \bar{\lambda}^{\rho \sigma},
\end{eqnarray}
here bar denotes formal complex conjugate $ \overline{\left(
\lambda^{\mu \alpha} \right)} = \bar{\lambda}^{\mu \dot{\alpha}}$.

\section{Properties of the Moyal-Weyl star product}
The Fourier transformation of chiral superfield $f(y,\theta)$ is
\begin{eqnarray}
& & \tilde{f}(k,\pi) = \int \! \frac{d^4 y}{\sqrt{(2\pi)^4}} \int \! 2 
d^2 \theta \ e^{i k^{\mu} y_{\mu} + i \pi^{\alpha} \theta_{\alpha}} 
f(y,\theta) ,\\
& & f(y,\theta) = \int \! \frac{d^4 k}{\sqrt{(2\pi)^4}} \int \! 2 
d^2 \pi \ e^{-i k^{\mu} y_{\mu} - i \pi^{\alpha} \theta_{\alpha}} 
\tilde{f}(k,\pi).
\end{eqnarray}
In our convention,
\begin{eqnarray}
& &\int \! d^2 \pi \ e^{i\pi^{\alpha} (\theta- \theta')_{\alpha}} \nonumber \\
& & = \int \! d^2 \pi \ \left( 1 + i \pi^{\alpha} (\theta - 
\theta)_{\alpha} + \frac{1}{4} \pi^2 ( \theta - \theta')^2 \right) 
\nonumber \\
& & = \frac{1}{4} \delta^{(2)} ( \theta - \theta').
\end{eqnarray}
The product of superfields on the noncommutative superspace
$[\hat{y}^{\mu} ,\hat{\theta}^{\alpha}] = i \lambda^{\mu \alpha}$ is 
\begin{eqnarray}
\hat{f}(\hat{y},\hat{\theta}) \hat{g}(\hat{y},\hat{\theta}) = \int \! 
\frac{d^4 y_1}{\sqrt{(2 \pi)^4}}  \int \! 
\frac{d^4 y_2}{\sqrt{(2 \pi)^4}} \int \! 2 d^2 \pi_1 \int \! 2 d^2 \pi_2 
e^{-ik_1^{\mu} \hat{y}_{\mu} - i \pi_1^{\alpha} 
\hat{\theta}_{\alpha}} e^{-ik_2^{\mu} \hat{y}_{\mu} - i \pi_2^{\alpha} 
\hat{\theta}_{\alpha}}  \tilde{f} (k_1,\pi_1) \tilde{g} (k_2,\pi_2).
\end{eqnarray}
We can use the BCH formula to combine the exponential sector,
\begin{eqnarray}
e^{-i (k_1+k_2)^{\mu} \hat{y}_{\mu} - i (\pi_1+\pi_2)^{\alpha} 
\hat{\theta}_{\alpha}+ i \Delta},
\end{eqnarray}
here the factor $i\Delta$ is
\begin{equation}
i \Delta = \frac{i}{2} \lambda_{\mu \alpha} \left( k_1^{\mu} 
\pi_2^{\alpha} - \pi_1^{\alpha} k_2^{\mu} \right).
\end{equation}
We then define the noncommutative star product between ordinary chiral superfields
\begin{eqnarray}
\!\! f(y,\theta) \star g(y,\theta) \equiv \int \!\! \frac{d^4 k_1}{\sqrt{(2 
\pi)^4}} \int \!\! \frac{d^4 k_2}{\sqrt{(2 
\pi)^4}} \int \!\! 2 d^2 \pi_1 \int \!\! 2 d^2 \pi_2 \ e^{- i (k_1 + 
k_2)^{\mu} y_{\mu} - i (\pi_1+\pi_2)^{\alpha} \theta_{\alpha}} 
e^{i\Delta} \tilde{f}(k_1,\pi_1) \tilde{g}(k_2,\pi_2), \label{Moyal-Weyl}
\end{eqnarray}
from which we can find explicit form of the Moyal-Weyl star product. The parameter transformation
\begin{eqnarray}
& & k_1 = \frac{k}{2} + k' , \quad k_2 = \frac{k}{2} - k' ,\nonumber \\
& & \pi_1 = \frac{\pi}{2} + \pi' , \quad \pi_2 = \frac{\pi}{2} - \pi',
\end{eqnarray}
gives Jacobian = 1 and 
\begin{eqnarray}
\widetilde{f\star g} (0) &=& \int \! \frac{d^4 k'}{\sqrt{(2\pi)^4}} \int \! 
2 d^2 \pi' \ \tilde{f}(k',\pi') \tilde{g}(-k',-\pi').
\end{eqnarray}
By taking $k=0$, the extra phase factor vanish, and we find
\begin{equation}
\int \! d^4 y d^2 \theta \ f(y,\theta) \star g(y,\theta) = \int \! d^4 y d^2 \theta \ f(y,\theta) g(y,\theta) .
\end{equation}
So, free part of the theory is not deformed by the star product.

Next, let us check the associativity $ f(y,\theta) \star
\left( g(y,\theta) \star h(y,\theta) \right) = \left( f(y,\theta) \star
g(y,\theta) \right) \star h(y,\theta) $. Fourier transformation of the chiral
superfields are
\begin{eqnarray}
f(y,\theta) &=& \int \! \frac{d^4 k}{\sqrt{(2 \pi)^4}} \int \! 2 d^2 \pi
 \ e^{-ik\cdot y - i \pi \cdot \theta} \tilde{f} (k,\pi) ,\nonumber \\
g(y,\theta) &=& \int \! \frac{d^4 p}{\sqrt{(2 \pi)^4}} \int \! 2 d^2 \rho
 \ e^{-ip\cdot y - i \rho \cdot \theta} \tilde{g} (p,\rho) ,\nonumber \\
h(y,\theta) &=& \int \! \frac{d^4 q}{\sqrt{(2 \pi)^4}} \int \! 2 d^2 \sigma
 \ e^{-iq\cdot y - i \sigma \cdot \theta} \tilde{h} (q,\sigma) .
\end{eqnarray}
It is enough to check
\begin{eqnarray}
e^{-ik\cdot y - i \pi \cdot \theta} \star \left( e^{-ip\cdot y - i \rho
					   \cdot \theta} \star
					   e^{-iq\cdot y - i \sigma
					   \cdot \theta} \right) =
\left( e^{-ik\cdot y - i \pi \cdot \theta} \star e^{-ip\cdot y - i \rho
 \cdot \theta} \right) \star e^{-iq\cdot y - i \sigma \cdot \theta}.
\end{eqnarray}
Calculation is straightforward by the use of BCH formula,
\begin{eqnarray}
\mathrm{LHS} &=& e^{-i(k+p+q) \cdot y - i (\pi + \rho + \sigma)\cdot
 \theta} e^{\frac{i}{2} \lambda_{\mu \alpha} \left[ k^{\mu} (\rho +
					      \sigma)^{\alpha} -
					      (p+q)^{\mu} \pi^{\alpha} +
					     p^{\mu} \sigma^{\alpha} -
					     q^{\mu} \rho^{\alpha}
					     \right]} ,\\
\mathrm{RHS} &=& e^{-i(k+p+q) \cdot y - i (\pi + \rho + \sigma)\cdot
 \theta} e^{\frac{i}{2} \lambda_{\mu \alpha} \left[ k^{\mu}
					      \rho^{\alpha} - p^{\mu}
					      \pi^{\alpha} + (k+p)^{\mu}
					     \sigma^{\alpha} - q^{\mu} (
					     \pi + \rho)^{\alpha}
					     \right]}.
\end{eqnarray}
It is obvious that $\mathrm{LHS} = \mathrm{RHS}$.

\end{appendix}

\end{document}